\documentclass[11pt]{amsart}

\usepackage{fullpage}
\usepackage[latin1]{inputenc}
\usepackage{bbm,natbib}
\usepackage{amsmath}
\usepackage{stmaryrd}
\usepackage{alltt, amssymb}
\usepackage{graphicx}
\usepackage{url}

\def\Jac{{\mathbf{Jac}}}

\newcommand{\sC}{\mathsf{C}}
\newcommand{\sL}{{\mathsf{L}}}
\newcommand{\sM}{{\mathsf{M}}}
\newcommand{\N}{\mathbb{N}}
\newcommand{\cO}{{\mathcal O}}
\newcommand{\order}{{r}}
\newcommand{\precision}{{N}}
\newcommand{\basefield}{{\mathbb{K}}}

\newcommand{\Mat}{{\mathsf{MM}}}
\renewcommand{\proof}{\noindent\textsc{Proof.} }
\newcommand{\foorp}{\hfill$\square$}

\newcommand{\trunc}[3]{\left[ #1 \right]_{#2}^{#3}}
\newcommand{\truncl}[2]{\left\lfloor #1 \right\rfloor_{#2}}
\newcommand{\trunch}[2]{\left\lceil #1 \right\rceil^{#2}}
\newcommand{\intpart}[1]{\left\lfloor #1 \right\rfloor}

\newtheorem{Theo}{Theorem}
\newtheorem{Prop}{Proposition}
\newtheorem{Lemme}{Lemma}

\usepackage{graphicx}
\usepackage{changebar}
\usepackage[plainpages=false,pdfpagelabels,colorlinks=true,citecolor=blue,hypertexnames=false]{hyperref}

\begin{document}

\title{Fast computation of power series solutions \\ of systems of
  differential equations}

\author{A. Bostan, F. Chyzak, F. Ollivier, B. Salvy, \'E. Schost, and A. Sedoglavic}
\thanks{Partially supported by a grant from the French \emph{Agence nationale pour la recherche}.}
%\date{Preliminary version 1.10 --- 11/04/2006}

\begin{abstract}
  We propose new algorithms for the computation of the first~$\precision$ terms
  of a vector (resp.\ a basis) of power series solutions of a linear
  system of differential equations at an ordinary point, using a
  number of arithmetic operations which is quasi-linear with respect
  to~$\precision$.  Similar results are also given in the non-linear case. This extends
  previous results obtained by Brent and
  Kung for scalar differential equations of order one and two.
\end{abstract}
\maketitle

%%%%%%%%%%%%%%%%%%%%%%%%%%%%%%%%%%%%%%%%%%%%%%%%%%%%%%%%%%%%
%%%%%%%%%%%%%%%%%%%%%%%%%%%%%%%%%%%%%%%%%%%%%%%%%%%%%%%%%%%%
%%%%%%%%%%%%%%%%%%%%%%%%%%%%%%%%%%%%%%%%%%%%%%%%%%%%%%%%%%%%

\section{Introduction}

In this article, we are interested in the computation of the first $\precision$ terms of power  
series solutions of differential equations. This problem arises in  
combinatorics, where the desired power series is a generating  
function, as well as in numerical analysis and in particular in  
control theory.

Let~$\basefield$ be a field. Given~$r+1$ formal power
series~${a_0(t),\dots,a_{\order}(t)}$ in~$\basefield[[t]]$, one of
our aims is to provide fast algorithms for solving the
linear differential equation
% of order $\order$:
\begin{equation} \label{lindiffeq}
a_\order(t) y^{(\order)}(t) + \dots + a_1(t) y'(t)+ a_0(t) y(t) = 0. %
\end{equation}
Specifically, under the hypothesis that~$t=0$ is an ordinary point
for Equation~\eqref{lindiffeq} (i.e., ${a_r(0) \neq 0}$), we give efficient
algorithms taking as input the first~$\precision$ terms of the power
series $a_0(t), \dots, a_\order(t)$ and answering the following algorithmic questions:
\begin{enumerate}
\item[{\bf i.}]  find the first~$\precision$ coefficients of
  the~$\order$ elements of a basis of power series solutions
  of~\eqref{lindiffeq};
\item[{\bf ii.}] given initial conditions~$\alpha_0, \dots,
  \alpha_{\order-1}$ in~$\basefield$, find the first~$\precision$
  coefficients of the unique solution~$y(t)$ in~$\basefield[[t]]$ of
  Equation~\eqref{lindiffeq} satisfying 
\[
  y(0) = \alpha_0,\quad y'(0) = \alpha_1, \quad \dots,\quad y^{(\order-1)}(0) =
  \alpha_{\order-1}.
\]
\end{enumerate}
More generally, we also treat linear first-order systems of differential
equations. From the data of initial conditions~$v$
in~$\mathcal{M}_{\order\times\order} (\basefield)$
(resp.~$\mathcal{M}_{{\order} \times 1} (\basefield)$) and of the
first~$\precision$ coefficients of each entry of the matrices~$A$
and~$B$ in~$\mathcal{M}_{\order\times\order} (\basefield[[t]])$ (resp.~$b$
in~$\mathcal{M}_{{\order} \times 1} (\basefield[[t]])$), we propose
algorithms that compute the first~$\precision$ coefficients:
\begin{enumerate}
\item[\bf I.]  of a fundamental solution~$Y$ in~$\mathcal{M}_{\order\times\order}
  (\basefield[[t]])$ of~${Y' = AY + B}$, with~${Y(0)=v},\;{\det Y(0) \neq 0}$;
\item[\bf II.]  of the unique solution~$y(t)$
  in~$\mathcal{M}_{{\order} \times 1} (\basefield[[t]])$ of~${y' = Ay
    + b}$, satisfying~${y(0) =v}$.
\end{enumerate}
%% \begin{equation}\label{systlindiffeq:basis}
%% Y' = AY + B, \quad \text{with} \; A,B \in \mathcal{M}_{\order} (\basefield[[t]])
%% \end{equation}
%% and 
%% \begin{equation}\label{systlindiffeq:single}
%% y' = Ay + b
%% \end{equation}
Obviously, if an algorithm of algebraic complexity~$\sC$ (i.e.,
using~$\sC$ arithmetic operations in~$\basefield$) is available for
problem~{\bf II}, then applying it~$r$ times solves problem~{\bf I} in
time~$r \,\sC$, while applying it to a companion matrix solves
problem~{\bf ii} in time~$\sC$ and problem~{\bf i} in~$r
\,\sC$. Conversely, an algorithm solving~{\bf i} (resp. {\bf I}) also
solves {\bf ii} (resp. {\bf II}) within the same complexity, plus that
of a linear combination of series. Our reason for distinguishing the
four problems {\bf i, ii, I, II} is that in many cases, we are able to
give algorithms of better complexity than obtained by these
reductions.

The most popular way of solving~{\bf i}, {\bf ii}, {\bf I}, and~{\bf II} is the
method of undetermined coefficients that requires~$\cO(\order^2
\precision^2)$ operations in~$\basefield$ for problem~{\bf i}
and~$\cO(\order \precision^2)$ operations in~$\basefield$ for~${\bf
  ii}$. Regarding the dependence in~$\precision$, this is certainly
too expensive compared to the size of the output, which is only linear
in~$\precision$ in both cases. On the other hand, verifying the
correctness of the output for~{\bf ii} (resp.~{\bf i}) already
requires a number of operations in~$\basefield$ which is linear
(resp.\ quadratic) in~$\order$: this indicates that there is little
hope of improving the dependence in~$\order$.  Similarly, for
problems~{\bf I} and~{\bf II}, the method of undetermined coefficients
requires~$\cO(\precision^2)$ multiplications of~$\order\times \order$
scalar matrices (resp.\ of scalar matrix-vector products in
size~$\order$), leading to a computational cost which is reasonable
with respect to~$\order$, but not with respect to~$\precision$.

By contrast, the algorithms proposed in this article have costs that
are linear (up to logarithmic factors) in the
complexity~$\sM(\precision)$ of polynomial multiplication in degree
less than~$\precision$ over~$\basefield$. Using Fast Fourier Transform
(FFT) these costs become nearly linear~---~up to polylogarithmic
factors~---~with respect to~$\precision$, for all of the four problems
above (precise complexity results are stated below).  Up to these
polylogarithmic terms in~$\precision$, this estimate is probably not
far from the lower algebraic complexity one can expect: indeed, the
mere check of the correctness of the output requires, in each case, a
computational effort proportional to~$\precision$.

%%%%%%%%%%%%%%%%%%%%%%%%%%%%%%%%%%%%%%%%%%%%%%%%%%%%%%%%%%%%

\subsection{Newton Iteration}
In the case of first-order equations ($r=1$), Brent and Kung have
shown in~\cite{BrKu78} (see also~\cite{Geddes1979,KuTr78}) that the problems
can be solved with complexity $\cO(\sM(\precision))$ by means of a
formal Newton iteration. Their algorithm is based on the fact that
solving the first-order differential equation~${y'(t) = a(t) y(t)}$,
with~$a(t)$ in~$\basefield[[t]]$ is equivalent to computing the
\emph{power series exponential\/}~$\exp(\int a(t))$.  This equivalence
is no longer true in the case of a system~${Y' = A(t) Y}$
(where~$A(t)$ is a power series matrix): for non-commutativity
reasons, the matrix exponential~${Y(t)= \exp(\int A(t))}$ is not a
solution of~${Y' = A(t) Y}$.

Brent and Kung suggest a way to extend their result to higher orders,
 and the corresponding algorithm has been shown by van der Hoeven
 in~\cite{vdHoeven02} to have complexity~$\cO(\order^\order
 \,\sM(\precision))$. This is good with respect to~$\precision$, but
 the exponential dependence in the order~$\order$ is unacceptable.

Instead, we solve this problem by devising a specific Newton iteration
for~${Y' = A(t) Y}$.  Thus we solve problems {\bf i} and {\bf I} in
$\cO(\Mat(\order,\precision))$, where $\Mat(\order,\precision)$ is the
number of operations in $\basefield$ required to multiply
$\order\times\order$ matrices with polynomial entries of degree less
than~$\precision$. For instance, when $\basefield=\mathbb{Q}$, this is
$\cO(\order^\omega \precision+r^2\sM(\precision))$, where
$\order^\omega$~can be seen as an abbreviation for~$\Mat(\order,1)$, see
\S\ref{ssec:complexity} below.

%%%%%%%%%%%%%%%%%%%%%%%%%%%%%%%%%%%%%%%%%%%%%%%%%%%%%%%%%%%%

\subsection{Divide-and-conquer}
The resolution of problems {\bf i} and {\bf I} by Newton iteration
relies on the fact that a whole basis is computed. Dealing with
problems {\bf ii} and {\bf II}, we do not know how to preserve this
algorithmic structure, while simultaneously saving a factor $\order$.

To solve problems~{\bf ii} and~{\bf II}, we therefore propose an
alternative algorithm, whose complexity is also nearly linear
in~$\precision$ (but not quite as good, being in
$\cO(\sM(\precision)\log\precision)$), but whose dependence in the
order~$\order$ is better~---~linear for~{\bf i} and quadratic for~{\bf
ii}. In a different model of computation with power series, based on
the so-called \emph{relaxed multiplication}, van der Hoeven briefly outlines
another algorithm~\cite[Section~4.5.2]{vdHoeven02} solving
problem~{\bf ii} in~$\cO(\order \,\sM(\precision) \log \precision)$.
To our knowledge, this result cannot be transferred to the usual model
of power series multiplication (called zealous in~\cite{vdHoeven02}).

We use a divide-and-conquer technique similar to that used in the fast
Euclidean algorithm~\cite{Knuth70,Schonhage71,Strassen83}. For
instance, to solve problem~{\bf ii}, our algorithm divides it into two
similar problems of halved size. The key point is that the lowest
coefficients of the solution~$y(t)$ only depend on the lowest
coefficients of the coefficients~$a_i$.  Our algorithm first computes
the desired solution~$y(t)$ at precision only~$\precision/2$, then it
recovers the remaining coefficients of~$y(t)$ by recursively solving
at precision~$\precision/2$ a new differential equation.  The main
idea of this second algorithm is close to that used for solving
first-order difference equations in~\cite{GaGe97}.

We encapsulate our main complexity results in
Theorem~\ref{theo:linear} below.  When FFT is used, the
functions~$\sM(\precision)$ and~$\Mat(\order,\precision)$ have, up to logarithmic terms, a nearly linear
growth in~$\precision$, see
\S\ref{ssec:complexity}. Thus, the results in the following theorem are quasi-optimal.
\begin{Theo}\label{theo:linear}
  Let~$\precision$ and~$\order$ be two positive integers and
  let\/~$\basefield$ be a field of characteristic zero or at
  least~$\precision$. Then:
  \begin{enumerate}
  \item[(a)] problems\/~{\bf i} and\/~{\bf I} can be solved
    using~$\cO\left(\Mat(\order,\precision) \right)$ operations
    in~$\basefield$;
  \item[(b)] problem\/~{\bf ii} can be solved using~$\cO\left(\order \,
    \sM (\precision) \log \precision\right)$ operations in~$\basefield$;
  \item[(c)] problem\/~{\bf II} can be solved using~$\cO\left(\order^2 \,
    \sM (\precision) \log \precision\right)$ operations in~$\basefield$.
  \end{enumerate}
\end{Theo}

%%%%%%%%%%%%%%%%%%%%%%%%%%%%%%%%%%%%%%%%%%%%%%%%%%%%%%%%%%%%
  
\subsection{Special Coefficients}  
For special classes of coefficients, we give different algorithms of
better complexity. We isolate two important classes of equations: that
with constant coefficients and that with polynomial coefficients.  In
the case of constant coefficients, our algorithms are based on the use
of the Laplace transform, which allows us to reduce the resolution of
differential equations with constant coefficients to manipulations
with rational functions.  The complexity results are summarized in the following theorem.
\begin{Theo}
  Let~$\precision$ and~$\order$ be two positive integers and
  let\/~$\basefield$ be a field of characteristic zero or at
  least~$\precision$. Then, for differential equations and systems with constant coefficients:
  \begin{enumerate}
  \item[(a)] problem\/~{\bf i} can be solved
    using~$\cO\left(\sM(\order)\,(\order+\precision) \right)$ operations
    in~$\basefield$;
  \item[(b)] problem\/~{\bf ii} can be solved using~$\cO\left(\sM(\order)\,(1+\precision/\order)\right)$ operations in~$\basefield$;
  \item[(c)] problem\/~{\bf I} can be solved using~$\cO\left( \order^{\omega+1}\log\order + \order\sM(\order)\precision \right)$ operations in~$\basefield$;
  \item[(d)] problem\/~{\bf II} can be solved using~$\cO\left( \order^\omega\log\order + \sM(\order)\precision \right)$ operations in~$\basefield$.
  \end{enumerate}
\end{Theo}
In the case of polynomial coefficients, we
exploit the linear recurrence satisfied by the coefficients of
solutions.  In Table~\ref{table1}, we gather the complexity estimates
corresponding to the best known solutions for each of the four
problems {\bf i}, {\bf ii}, {\bf I}, and~{\bf II} in the general case,
as well as in the above mentioned special cases. The algorithms are described in Section~\ref{sec:particular}.  In the polynomial
coefficients case, these results are well known. In the other cases,
to the best of our knowledge, the results improve upon existing
algorithms.

\begin{table}
\renewcommand{\arraystretch}{1.4}
$$\begin{array}{||l|l|l|l||l||}\hline\hline   % & & & & \\
 \textsf{Problem} & \textsf{constant} & \textsf{polynomial}
& \textsf{power series} & \textsf{output}\\[-2mm]
\quad (\textsf{input, output}) & \textsf{coefficients} &
\textsf{coefficients} & \textsf{coefficients} & \textsf{size} \\  
% & & & & \\

\hline \hline \textbf{i} \quad  (\textsf{equation, basis}) &   \cO(\sM(\order)
 \precision) \;\hfill^\star & \cO(d \order^2 \precision) &   \cO(
 \Mat(\order, \precision)) \;\hfill ^\star &  \cO(\order \precision)\\

\hline  \textbf{ii} \quad  (\textsf{equation, one solution}) &
 \cO(\sM(\order)  \precision/\order) \;\hfill^\star  &\cO(d
 \order \precision)  &\cO(\order \, \sM(\precision) \log \precision) \;\hfill ^\star & \cO(\precision)\\

 \hline \hline
 \textbf{I} \quad (\textsf{system, basis}) &
 \cO(\order \sM(\order)
 \precision) \;\hfill^\star &  \cO(d \order^\omega \precision) 
 & \cO(\Mat(\order, \precision))  \;\hfill ^\star  & \cO(\order^2 \precision)\\

\hline \textbf{II} \quad  (\textsf{system, one solution})  &
 \cO(\sM(\order) \precision) \;\hfill ^\star &  \cO(d \order^2
 \precision)  & \cO(\order^2 \, \sM(\precision) \log
 \precision) \;\hfill ^\star  & \cO(\order \precision)\\
 
\hline\hline
%\quad \quad \textsf{Input size} & \cO(\order^2)  &  \cO(d \order^2)  & \cO(\order^2
% \precision) \\ \hline \hline  
\end{array}$$
\caption{Complexity of solving linear differential equations/systems for~$\precision\gg\order$.  Entries marked with a~`$\star$' correspond to new results. \label{table1}}
\end{table}

%%%%%%%%%%%%%%%%%%%%%%%%%%%%%%%%%%%%%%%%%%%%%%%%%%%%%%%%%%%%

\subsection{Non-linear Systems}  As an important
consequence of Theorem~\ref{theo:linear}, we improve the known
complexity results for the more general problem of solving
\emph{non-linear} systems of differential equations. To do so, we use
a classical reduction technique from the non-linear to the linear
case, see for instance~\cite[Section~25]{Rall69}
and~\cite[Section~5.2]{BrKu78}. For simplicity, we only consider
non-linear systems of first order. There is no loss of generality in
doing so, more general cases can be reduced to that one by adding new
unknowns and possibly differentiating once. The following result
generalizes~\cite[Theorem~5.1]{BrKu78}.  If~${F=(F_1,\dots,F_r)}$ is a
differentiable function bearing on~$\order$
variables~${y_{1},\dots,y_{\order}}$, we use the notation~$\Jac(F)$
for the Jacobian matrix~$(\partial F_i/\partial y_j)_{1\leq i,j \leq \order}$.

\begin{Theo}\label{theo:non-linear}
  Let~$\precision$, $\order$ be in~$\mathbb{N}$, let~$\basefield$ be a
  field of characteristic zero or at least\/~$\precision$ and
  let~$\varphi$ denote~${(\varphi_1,\dots,\varphi_{\order})}$,
  where~$\varphi_i(t,y)$ are multivariate power series
  in\/~$\basefield[[t,y_1,\dots,y_\order]]$.
  \par
  Let\/~${\sL :\N \to \N}$ be such that for all~$s(t)$
  in\/~$\mathcal{M}_{\order \times 1}(\basefield[[t]])$ and for all~$n$
  in\/~$\mathbb{N}$, the first~$n$ terms of~$\varphi(t,s(t))$ and
  of\/~$\Jac ({\varphi}) (t,s(t))$ can be computed in~$\sL(n)$ operations
  in\/~$\basefield$.  Suppose in addition that the function~${n \mapsto
    \sL(n)/n}$ is increasing. Given initial conditions~$v$
  in\/~$\mathcal{M}_{\order \times 1}(\basefield)$, if the differential
  system
  \[y'=\varphi(t,y),\qquad y(0)=v,\] admits a solution 
  in\/~$\mathcal{M}_{\order \times 1} (\basefield[[t]])$, then the 
first\/~$\precision$ terms of such a solution~$y(t)$ can be computed in
%  $\cO(\sL(N) + \Mat(\order,\precision))$ operations in $\basefield$.
%  $\cO(\sL(N) + \order^2 \sM(\precision) \log \precision)$ operations in $\basefield$.
  $\cO \left(\sL(\precision) + \min (\Mat(\order,\precision), \order^2
  \sM(\precision) \log \precision) \right)$ operations in~$\basefield$.
\end{Theo}
Werschulz~\cite[Theorem~3.2]{Werschulz80} gave an algorithm solving
the same problem using the integral Volterra-type equation technique
described in~\cite[pp.~172--173]{Rall69}.  With our notation, his
algorithm uses~$\cO \left(\sL(\precision) + \order^2 \precision \,
\sM(\precision)) \right)$ operations in~$\basefield$ to compute a
solution at precision~$\precision$. Thus, our algorithm is an
improvement for cases where $\sL(\precision)$ is known to be
subquadratic with respect to~$\precision$.

The best known algorithms for power series composition in~${\order
\geq 2}$ variables require, at least on ``generic'' entries, a
number~${\sL(n) = \cO(n^{\order-1} \sM(n))}$ of operations in
$\basefield$ to compute the first~$n$ coefficients of the
composition~\cite[Section~3]{BrKu77}.  This complexity is nearly
optimal with respect to the size of a generic input. By contrast, in
the univariate case, the best known result~\cite[Th.~2.2]{BrKu78}
is~$\sL(n) = \cO(\sqrt{n \log n}\, \sM(n))$. For special entries,
however, better results can be obtained, already in the univariate
case: exponentials, logarithms, powers of univariate power series can
be computed~\cite[Section~13]{Brent75} in~$\sL(n) = \cO(\sM(n))$. As a
consequence, if~$\varphi$ is an~$\order$-variate sparse polynomial
with $m$~monomials of \emph{any} degree, then~$\sL(n) = \cO(m \order \,
\sM(n))$.

Another important class of systems with such a
subquadratic~$\sL(\precision)$ is provided by \emph{rational systems},
where each~$\varphi_i$ is in~$\basefield(y_1,\dots,y_\order)$.
Supposing that the complexity of evaluation of~$\varphi$ is bounded
by~$L$ (i.e., for any point~$z$ in~$\basefield^\order$ at
which~$\varphi$ is well-defined, the value~$\varphi(z)$ can be
computed using at most~$L$ operations in~$\basefield$), then, the
Baur-Strassen theorem~\cite{BaSt83} implies that the complexity of
evaluation of the Jacobian~$\Jac(\varphi)$ is bounded by~$5L$, and
therefore, we can take~${\sL(n)= \sM(n) L}$ in the statement of
Theorem~\ref{theo:non-linear}.

%%%%%%%%%%%%%%%%%%%%%%%%%%%%%%%%%%%%%%%%%%%%%%%%%%%%%%%%%%%%

\subsection{Basic Complexity Notation} \label{ssec:complexity}

Our algorithms ultimately use, as a basic operation, multiplication
of matrices with entries that are polynomials (or truncated power
series).  Thus, to estimate their complexities in a unified manner,
we use a function~${\Mat : \N \times \N \to \N}$ such that any two~${r
  \times r}$ matrices with polynomial entries in~$\basefield[t]$ of
degree less than~$d$ can be multiplied using~$\Mat(r,d)$ operations
in~$\basefield$. In particular,~$\Mat(1,d)$ represents the number of
base field operations required to multiply two polynomials of degree
less than~$d$, while~$\Mat(r,1)$ is the arithmetic cost of scalar~${r
  \times r}$ matrix multiplication. For simplicity, we
denote~$\Mat(1,d)$ by~$\sM(d)$ and we have~${\Mat(r,1) =
  \cO(r^\omega)}$, where~${2 \leq \omega \leq 3}$ is the so-called {\em
  exponent of the matrix multiplication}, see, e.g.,~\cite{BuClSh97}
and~\cite{GaGe99}.

Using the algorithms of~\cite{ScSt71,CaKa91}, one can take~$\sM(d)$
in~$\cO(d \log d \log \log d)$; over fields supporting FFT, one can
take~$\sM(d)$ in~$\cO(d\log d)$.  By~\cite{CaKa91} we can always
choose~$\Mat(r,d)$ in~${\cO(r^\omega \, \sM(d))}$, but better
estimates are known in important particular cases.  For instance, over
fields of characteristic~$0$ or larger than~$2d$, we have~${\Mat(r,d)
  = \cO( r^\omega d + r^2 \, \sM(d))}$, see~\cite[Th.~4]{BoSc05}.  To
simplify the complexity analyses of our algorithms, we suppose that the
{multiplication cost} function~$\Mat$ satisfies the following standard
growth hypotheses for all integers~$d_{1},d_{2}$ and~$r$: %(see, e.g., \cite{GaGe99}).
\begin{equation}\label{hyp:Mat}
\Mat(r,d_{1}d_{2}) \leq d_{1}^{2} \Mat (r,d_{2})
\qquad \text{and}  \qquad
\frac{\Mat(r,d_{1})}{d_{1}} \leq \frac{\Mat(r,d_{2})}{d_{2}}
\quad  \text{if $d_{1} \leq d_{2}$}.
\end{equation}
In particular, Equation~\eqref{hyp:Mat} implies the inequalities
\begin{equation} \label{ineq:Mat}
\begin{split}
	\Mat(r,2^\kappa)+\Mat(r,2^{\kappa-1})+M(r,2^{\kappa-2})+\dots+\Mat(r,1)&
		\le 2\Mat(r,2^\kappa),\\
	\sM(2^\kappa)+2\sM(2^{\kappa-1})+4\sM(2^{\kappa-2})+\dots+2^\kappa\sM(1)&
		\le (\kappa+1)\sM(2^\kappa).
\end{split}
\end{equation}
These inequalities are crucial to prove the estimates in
Theorem~\ref{theo:linear} and Theorem~\ref{theo:non-linear}.  Note
also that when the available multiplication algorithm is slower than
quasi-linear (e.g., Karatsuba or naive multiplication), then in the
second inequality, the factor~$(\kappa+1)$ can be replaced by a constant
and thus the estimates $\sM(\precision)\log \precision$ in our complexities become
$\sM(\precision)$ in those cases.

%%%%%%%%%%%%%%%%%%%%%%%%%%%%%%%%%%%%%%%%%%%%%%%%%%%%%%%%%%%%

\subsection{Notation for Truncation}

It is recurrent in algorithms to split a polynomial into a lower and a
higher part. To this end, the following notation proves convenient.
Given a polynomial~$f$, the remainder and quotient of its Euclidean
division by~$t^k$ are respectively denoted $\trunch fk$ and~$\truncl
fk$.  Another occasional operation consists in taking a middle part
out of a polynomial.  To this end, we let $\trunc fkl$
denote~$\truncl{\trunch fl}{k}$.  Furthermore, we shall write $f=g\mod
t^k$ when two polynomials or series $f$ and~$g$ agree up to
degree~$k-1$ included.  To get a nice behaviour of integration with
respect to truncation orders, all primitives of series are chosen with
zero as their constant coefficient.

%%%%%%%%%%%%%%%%%%%%%%%%%%%%%%%%%%%%%%%%%%%%%%%%%%%%%%%%%%%%
%%%%%%%%%%%%%%%%%%%%%%%%%%%%%%%%%%%%%%%%%%%%%%%%%%%%%%%%%%%%
%%%%%%%%%%%%%%%%%%%%%%%%%%%%%%%%%%%%%%%%%%%%%%%%%%%%%%%%%%%%

\section{Newton Iteration for Systems of Linear Differential
  Equations}

Let~${Y'(t) = A(t) Y(t)+B(t)}$ be a linear differential system,
where~$A(t)$ and~$B(t)$ are~${\order \times \order}$ matrices with
coefficients in~$\basefield[[t]]$. Given an invertible scalar
matrix~$Y_0$, an integer~${\precision \geq 1}$, and the expansions
of~$A$ and~$B$ up to precision~$\precision$, we show in this section
how to compute efficiently the power series expansion at
precision~$\precision$ of the unique solution of the Cauchy problem
$$Y'(t) = A(t) Y(t)+B(t) \quad \text{and} \quad Y(0) = Y_0.$$ 
This enables us to answer problems \textbf{I} and \textbf{i}, the
latter being a particular case of the former (through the application
to a companion matrix).

%%%%%%%%%%%%%%%%%%%%%%%%%%%%%%%%%%%%%%%%%%%%%%%%%%%%%%%%%%%%

\subsection{Homogeneous Case}
First, we design a Newton-type iteration to solve the homogeneous
system~${Y'=A(t)Y}$.  The classical Newton iteration to solve an
equation $\phi(y)=0$ is $Y_{\kappa+1}=Y_\kappa-U_\kappa$, where
$U_\kappa$ is a solution of the linearized equation
$D\phi|_{Y_\kappa}\cdot U=\phi(Y_\kappa)$ and $D\phi|_{Y_\kappa}$ is
the differential of~$\phi$ at~$Y_\kappa$. We apply this idea to the
map~${\phi: Y \mapsto Y'-AY}$. Since~$\phi$ is linear, it is its own
differential and the equation for~$U$ becomes
$$U'-AU=Y'_\kappa-AY_\kappa.$$
Taking into account the proper orders of truncation and using
Lagrange's method of variation of
parameters~\cite{Lagrange1869,Ince56}, we are thus led to the
iteration
\[\begin{cases}Y_{\kappa+1} &= Y_\kappa - \trunch {U_\kappa}
{2^{\kappa+1}},\\
U_{\kappa}& = Y_\kappa \int
\trunch{Y_\kappa^{-1}}{2^{\kappa+1}} \left(Y_\kappa' -
  \trunch{A}{2^{\kappa+1}} Y_\kappa\right).
\end{cases}
\]
Thus we need to compute (approximations
of) the solution~$Y$ and its inverse simultaneously.  Now, a well-known Newton
iteration for the inverse $Z$ of $Y$ is
\begin{equation}\label{Newton:inverse}
Z_{\kappa+1} =
\trunch
  {Z_{\kappa} + Z_{\kappa} (I_\order - Y Z_{\kappa})}
  {2^{\kappa+1}}.
\end{equation}
 It was introduced by Schulz~\cite{Schulz33} in
the case of real matrices; its version for matrices of power series is
given for instance in~\cite{MoCa79}.

\begin{figure}
  \begin{center} 
    \fbox{\begin{minipage}{9cm}
      \medskip
      \begin{center}\textsf{SolveHomDiffSys}($A,\precision,Y_0$) \end{center}
      \textbf{Input:} ${Y_0,A_0, \dots, A_{\precision-2}}$
        in~$\mathcal{M}_{\order\times\order}(\basefield)$,
        ${A = \sum A_i t^i}$.
        \par\smallskip
        \textbf{Output:} ${Y=\sum_{i=0}^{\precision-1}Y_i t^i}$ in
$\mathcal{M}_{\order\times\order}(\basefield)[t]$ such that 
${Y' = A Y \mod t^{\precision-1}}$, and $Z=Y^{-1}\mod t^{\precision/2}$.

        \begin{tabbing}
          \;\;\\$Y \leftarrow  (I_{\order}+ t  A_0) Y_0$ 
          \\$ Z \leftarrow Y_0^{-1}$
          \\$m \leftarrow 2$\\
          \textsf{while} $m \leq \precision/2$ \textsf{do}\\
          \hspace{0.5cm} $Z \leftarrow Z + \trunch {Z(I_{\order} - YZ)}{m} $\\
          \hspace{0.5cm} $Y \leftarrow Y - \trunch {Y\left(\int Z (Y' - \trunch{A}{2m-1} Y) \right)}{2m} $ \\
                                %$+\sum_{i}\textsf{Coeff}(M', i)\frac{T^i}{i}$\\
          \hspace{0.5cm} $m \leftarrow 2m$ \\
          \textsf{return} $Y,Z$
        \end{tabbing}
      \end{minipage}
    }\end{center}
  \caption{Solving the Cauchy problem~$Y' = A(t) Y$,  $Y(0) = Y_0$ by Newton iteration.}
  \label{fig:hom}
\end{figure} 
Putting together these considerations, we arrive at the algorithm
\textsf{SolveHomDiffSys} in Figure~\ref{fig:hom}, whose correctness
easily follows from Lemma~\ref{prop:Newton} below.  Remark
that in the scalar case~(${\order=1}$) algorithm
\textsf{SolveHomDiffSys} coincides with the algorithm for power series
exponential proposed by Hanrot and Zimmermann~\cite{HaZi04}; see
also~\cite{Bernstein}. In the case~${\order>1}$, ours is a nontrivial
generalization of the latter. Because it takes primitives of series at
precision~$\precision$, algorithm \textsf{SolveHomDiffSys} requires
that the elements~${2,3,\dots,\precision-1}$ be invertible
in~$\basefield$. Its complexity~$\sC$ satisfies the
recurrence~${\sC(m) = \sC(m/2) + \cO(\sM(\order,m))}$, which
implies~---~using the growth hypotheses on~$\sM$~---~that~${\sC(\precision)
  = \cO(\sM(\order,\precision))}$.  This proves the first assertion of 
Theorem~\ref{theo:linear}.
%   It computes simultaneously the solutions $(Y,Z)$ of the problems
%   $$Y'-AY = 0 \bmod t^{\precision-1} \quad \text{and} \quad Z'+Z A = 0 \bmod
%   t^{\precision/2-1}.$$
%\bigskip

% This is based on the following result, allowing to double the
% precision of the solution, by using only polynomial matrix operations.
\smallskip

 \begin{Lemme}\label{prop:Newton} 
   Let~$m$ be an even integer. Suppose
   that~$Y_{(0)}, Z_{(0)}$ in~$\mathcal{M}_{\order\times\order}(\basefield[t])$ satisfy
   \begin{equation*}
     I_{\order} - Y_{(0)} Z_{(0)} = 0 \mod t^{m/2} \quad \text{and} \quad
     Y_{(0)}' - AY_{(0)} = 0 \mod t^{m-1},
   \end{equation*}
and that they are of degree less than $m/2$ and~$m$, respectively.
   Define
   \begin{equation*}
     Z:=\trunch {Z_{(0)} \left(2I_{\order}  - Y_{(0)} Z_{(0)} \right)} {m} \quad \text{and} \quad
     Y:=\trunch {Y_{(0)} \left(I_{\order} - \int Z  (Y_{(0)}'-AY_{(0)})  \right)} {2m}.
   \end{equation*}
   Then~$Y$ and~$Z$ satisfy the equations
   \begin{equation} \label{eq:double}
     I_{\order} - Y Z = 0 \mod t^{m} \quad \text{and} \quad
     Y' - AY = 0 \mod t^{2m-1}.
   \end{equation}
 \end{Lemme}
\proof Using the definitions of~$Y$ and~$Z$, it follows that
$$
I_{\order} - YZ = (I_{\order} -Y_{(0)} Z_{(0)})^2 - (Y -
Y_{(0)}) Z_{(0)} (2I_{\order} -Y_{(0)} Z_{(0)}) \mod t^m.
$$
Since by hypothesis~${I_{\order} -Y_{(0)} Z_{(0)}}$ and~${Y -
  Y_{(0)}}$ are zero modulo~$t^{m/2}$, the right-hand side is zero
modulo~$t^m$ and this establishes the first formula in
Equation~\eqref{eq:double}.  Similarly, write~${Q= \int Z
  (Y_{(0)}'-AY_{(0)})}$ and observe $Q=0\mod t^m$ to get the equality
$$
Y' - AY = (I-YZ) (Y_{(0)}' - AY_{(0)}) - (Y_{(0)}' -
AY_{(0)}) Q \mod t^{2m-1}.
$$
Now,~${Y_{(0)}' - AY_{(0)}}=0 \mod t^{m-1}$, while~$Q$
and~${I_{\order} -YZ}$ are zero modulo~$t^{m}$ and therefore
the right-hand side of the last equation is zero modulo~$t^{2m-1}$,
proving the last part of the lemma. 
\foorp

%%%%%%%%%%%%%%%%%%%%%%%%%%%%%%%%%%%%%%%%%%%%%%%%%%%%%%%%%%%%

\subsection{General Case}
We want to solve the equation~${Y'=AY +B}$, where~$B$ is an~${\order
\times \order}$ matrix with coefficients in~$\basefield[[t]]$.
Suppose that we have already computed the solution~$\widetilde{Y}$ of
the associate homogeneous equation~${\widetilde{Y}'=A \widetilde{Y}}$,
together with its inverse~$\widetilde{Z}$.  Then, by the method of
variation of parameters, ${Y_{(1)}= \widetilde{Y} \int \widetilde{Z}
B}$ is a particular solution of the inhomogeneous problem, thus the
general solution has the form~${Y = Y_{(1)}+\widetilde{Y}}$.

\begin{figure}
  \begin{center} 
    \fbox{\begin{minipage}{9.5cm}
        \medskip
        \begin{center}\textsf{SolveInhomDiffSys}($A,B,\precision,Y_0$) \end{center}
        \textbf{Input:} ${Y_0,A_0, \dots, A_{\precision-2}}$ in~$\mathcal{M}_{\order\times\order}(\basefield)$,
          ${A = \sum A_i t^i}$,
          \par\smallskip
          ${B_0, \dots, B_{\precision-2}}$ in~$\mathcal{M}_{\order\times\order}(\basefield)$, 
            ${B(t) = \sum B_i t^i}$.
            \par\medskip
            \textbf{Output:} ${Y_1,\dots,Y_{\precision-1}}$
            in~$\mathcal{M}_{\order\times\order}(\basefield)$ such that ${Y=Y_0 + \sum Y_i
            t^i}$ satisfies~${Y' = A Y + B \mod t^{\precision-1}}$.

\begin{tabbing}
  \;\;\\$\widetilde{Y},\widetilde{Z} \leftarrow \textsf{SolveHomDiffSys} (A,\precision,Y_0)$ \\
  $\widetilde{Z} \leftarrow \widetilde{Z} + \trunch {\widetilde{Z}(I_\order - \widetilde{Y}\widetilde{Z})} {\precision}$\\
  $Y \leftarrow \trunch {\widetilde{Y}   \int (\widetilde{Z}  B)} {\precision}$ \\
  $Y \leftarrow Y + \widetilde{Y}$\\
  \textsf{return} $Y$
\end{tabbing}
\end{minipage}
}\end{center}
\caption{Solving the Cauchy problem $Y' = A Y  + B, \; Y(0) = Y_0$ by Newton iteration.}
\label{fig:inhom}
\end{figure} 

Now, to compute the particular solution~$Y_{(1)}$ at
precision~$\precision$, we need to know both~$\widetilde{Y}$
and~$\widetilde{Z}$ at the same precision~$\precision$. To do this, we
first apply the algorithm for the homogeneous case and
iterate~\eqref{Newton:inverse} once. The resulting algorithm is
encapsulated in Figure~\ref{fig:inhom}.

%%%%%%%%%%%%%%%%%%%%%%%%%%%%%%%%%%%%%%%%%%%%%%%%%%%%%%%%%%%%
%%%%%%%%%%%%%%%%%%%%%%%%%%%%%%%%%%%%%%%%%%%%%%%%%%%%%%%%%%%%
%%%%%%%%%%%%%%%%%%%%%%%%%%%%%%%%%%%%%%%%%%%%%%%%%%%%%%%%%%%%

\section{Divide-and-conquer Algorithm}\label{sec:DAC}

We now give our second algorithm, which allows us to solve problems
{\bf ii} and~{\bf II} and to finish the proof of
Theorem~\ref{theo:linear}.  Before entering a detailed presentation,
let us briefly sketch the main idea in the particular case of a
homogeneous differential equation~${\mathcal{L}y=0}$,
where~$\mathcal{L}$ is a linear differential operator in~${\delta = t
\frac{d}{dt}}$ with coefficients in~$\basefield[[t]]$.
% FC, 11/04/2006: Je re-coupe, sinon cette phrase est trop longue !
(The introduction of~$\delta$ is only for pedagogical reasons.)  The
starting remark is that if a power series~$y$ is written as~${y_0 +
t^m y_1}$, then~${\mathcal{L}(\delta)y = \mathcal{L}(\delta)y_0 +
t^m\mathcal{L}(\delta + m)y_1}$. Thus, to compute a solution~$y$
of~${\mathcal{L}(\delta) y = 0 \mod t^{2m}}$, it suffices to determine
the lower part of~$y$ as a solution of ${\mathcal{L}(\delta) y_0 = 0
\mod t^m}$, and then to compute the higher part~$y_1$, as a solution
of the inhomogeneous equation~${\mathcal{L}(\delta + m) y_1 = - R \mod
t^{m}}$, where the rest~$R$ is computed so that~${\mathcal{L}(\delta)
y_0 = t^m R \mod t^{2m}}$.

Our algorithm \textsf{DivideAndConquer} makes a recursive use of this idea. Since, during the
recursions, we are naturally led to treat inhomogeneous equations of a
slightly more general form than that of~{\bf II} we introduce the
notation~$\mathcal{E}(s,p,m)$ for the vector equation
\begin{equation*}
t y' +   (p I_\order - tA) y =  s \mod t^{m}.
\end{equation*}
The algorithm is described in Figure~\ref{fig:algo-dac}.
Choosing~${p=0}$ and~${s(t) =t b(t)}$ we retrieve the equation of
problem~{\bf II}.  Our algorithm \textsf{Solve} to solve problem~{\bf
II} is thus a specialization of \textsf{DivideAndConquer}, defined by
making \textsf{Solve}$(A,b,\precision,v)$ simply call
\textsf{DivideAndConquer}$(tA,tb,0,\precision,v)$. Its correctness relies on
the following immediate lemma. 

\begin{figure}\label{fig:algo-dac}
  \begin{center} 
    \fbox{\begin{minipage}{8.5 cm}
        \medskip
        \begin{center}\textsf{DivideAndConquer($A,s,p,m,v$)} \end{center}
	\textbf{Input:} $A_0,\dots,A_{m-1}$ in~$\mathcal{M}_{\order\times\order}(\basefield)$,
        ${A = \sum A_i t^i}$, $s_0,\dots,s_{m-1},v$ in~$\mathcal{M}_{\order\times1}(\basefield)$,
	${s = \sum s_i t^i}$, $p$ in~$\basefield$.
        \par\smallskip
        \textbf{Output:} ${y=\sum_{i=0}^{\precision-1}y_i t^i}$ in
$\mathcal{M}_{\order\times1}(\basefield)[t]$ such that 
${ty' + (pI_{\order}-tA)y=s \mod t^m}$, ${y(0)=v}$.

        \begin{tabbing}
          \textsf{If}~$m=1$ \textsf{then} \\
          {\quad \textsf{if}} $p=0$ \textsf{then} \\
          {\quad \quad \textsf{return}} $v$\\
          {\quad else}  \textsf{return} $p^{-1} s(0)$\\
          \textsf{end if}\\
          $d \leftarrow \intpart{m/2}$\\
          $s \leftarrow \trunch s{d}$\\
          $y_0 \leftarrow$ {\sf DivideAndConquer}($A,s,p,d,v$)\\
          $R \leftarrow \trunc{s- t y_0' - (p I_\order -tA) y_0}{d}{m} $ \\
          $y_1 \leftarrow$ {\sf DivideAndConquer}($A, R, p+d, m-d,v$)\\
          \textsf{return} $y_0 + t^d y_1$
        \end{tabbing}
      \end{minipage}
    }\end{center}
  \caption{Solving $ty' + (pI_{\order}-tA)y=s \mod t^m$, ${y(0)=v}$, 
    by divide-and-conquer.}
\label{fig:2}
 \end{figure} 

\begin{Lemme}
  Let~$A$ in~$\mathcal{M}_{\order\times\order}(\basefield[[t]])$, $s$
  in~$\mathcal{M}_{\order \times 1}(\basefield[[t]])$, and let~$p,d$
  in~$\mathbb{N}$.  Decompose~$\trunch sm$ into a sum~${s_0 +
    t^d s_1}$.  Suppose that~$y_0$
  in~$\mathcal{M}_{\order\times1}(\basefield[[t]])$ satisfies the
  equation~$\mathcal{E}(s_0,p,d)$, set $R$ to be
  \begin{equation*}
    \trunch {(ty'_0 + (pI_\order - t A) y_0 - s_0)/t^d} {m-d},
  \end{equation*}
  and let~$y_1$ in~$\mathcal{M}_{\order \times 1}(\basefield[[t]])$ be
  a solution of the equation~${\mathcal{E}(s_1-R,p+d,m-d)}$.  Then the
  sum $y:= y_0 + t^d y_1$ is a solution of the
  equation~$\mathcal{E}(s,p,m)$.
\end{Lemme}

The only divisions performed along our algorithm~\textsf{Solve} are by 1, \dots, $\precision-1$.
As a consequence of this remark and of the previous lemma, we deduce the complexity estimates in the proposition below;
for a general matrix~$A$, this proves point~(c) in Theorem~\ref{theo:linear}, while the
particular case when $A$~is companion proves point~(b).

\begin{Prop}
  Given the first~$m$ terms of the entries
  of~$A\in\mathcal{M}_{\order\times\order}(\basefield[[t]])$ and
  of~$s\in\mathcal{M}_{\order \times 1}(\basefield[[t]])$,
  given~$v\in\mathcal{M}_{\order \times 1}(\basefield)$,
  algorithm~$\emph{\textsf{DivideAndConquer}}(A,s,p,m,v)$ computes a
  solution of the linear differential system~${ty' + (pI_{\order}-tA)
  y=s \mod t^m}$, ${y(0)=v}$, using~${\cO(\order^2 \, \sM(m) \log m)}$
  operations in~$\basefield$. If $A$ is a companion matrix, the cost 
  reduces to ${\sC(m) = \cO(\order \, \sM(m) \log m)}$.
\end{Prop}
\proof The correctness of the algorithm follows from the previous
Lemma.  The cost~$\sC(m)$ of the algorithm satisfies the recurrence
$$ \sC(m) = \sC(\intpart{m/2}) + \sC(\trunch{m/2}{}) + \order^2 \, \sM(m)
+ \cO(\order m),$$ where the term $\order^2 \, \sM(m)$ comes from the
application of $A$ to $y_0$ used to compute the rest~$R$. From this
recurrence, it is easy to infer that~${\sC(m) = \cO(\order^2 \, \sM(m)
\log m)}$. Finally, when $A$ is a companion matrix, the vector~$R$ can
be computed in time $O(\order \, \sM(m))$, which implies that in this
case~${\sC(m) = \cO(\order \, \sM(m) \log m)}$.
\foorp

%%%%%%%%%%%%%%%%%%%%%%%%%%%%%%%%%%%%%%%%%%%%%%%%%%%%%%%%%%%%
%%%%%%%%%%%%%%%%%%%%%%%%%%%%%%%%%%%%%%%%%%%%%%%%%%%%%%%%%%%%
%%%%%%%%%%%%%%%%%%%%%%%%%%%%%%%%%%%%%%%%%%%%%%%%%%%%%%%%%%%%

\section{Faster Algorithms for Special Coefficients}\label{sec:particular}

%%%%%%%%%%%%%%%%%%%%%%%%%%%%%%%%%%%%%%%%%%%%%%%%%%%%%%%%%%%%

\subsection{Constant Coefficients}\label{ssec:const-coeffs}
Let~$A$ be a constant~${\order \times \order}$ matrix and let~$v$ be a
vector of initial conditions. Given~${\precision \geq 1}$, we want to
compute the first~$\precision$ coefficients of the series expansion of
a solution~$y$ in~$\mathcal{M}_{\order \times 1}(\basefield[[t]])$
of~${y' = Ay}$, with~${y(0) = v}$. In this setting, many various
algorithms have been proposed to solve problems {\bf i}, {\bf ii},
{\bf I}, and {\bf II}, see for
instance~\cite{Pennell26,Putzer66,Kirchner67,Fulmer75,MoLo78,Leonard96,Liz98,Gu99,Gu01,HaFiSm01,MoLo03,LuRo04}.
Again, the most naive algorithm is based on the method of undetermined
coefficients. On the other hand, most books on differential equations,
see, e.g., \cite{Ince56,Coddington61,Arnold92} recommend to simplify
the calculations using the Jordan form of matrices. The main drawback
of that approach is that computations are done over the algebraic
closure of the base field~$\basefield$. The best complexity result
known to us is given in~\cite{LuRo04} and it is quadratic in~$\order$.

We concentrate first on problems~{\bf ii} and~{\bf II} (computing a
single solution for a single equation, or a first-order system).
Our algorithm for problem~{\bf II}
uses~${\cO(\order^\omega \log \order + \precision \sM(\order))}$
operations in~$\basefield$ for a general constant matrix~$A$ and
only~$\cO(\precision \sM(\order)/\order)$ operations in~$\basefield$ in
the case where~$A$ is a companion matrix (problem {\bf ii}). Despite
the simplicity of the solution, this is, to the best of our knowledge,
a new result.

In order to compute~${y_\precision = \sum_{i=0}^\precision{A^i v
t^i/i!}}$, we first compute its Laplace
transform~${z_\precision=\sum_{i=0}^\precision {A^i v t^i}}$: indeed,
one can switch from~$y_\precision$ to~$z_\precision$ using
only~$\cO(\precision \order)$ operations in $\basefield$.  The
vector~$z_\precision$ is the truncation at order~${\precision + 1}$
of~${z=\sum_{i\ge0} A^i v t^i =(I-tA)^{-1} v}$. As a byproduct of a
more difficult question,~\cite[Prop.~10]{Storjohann02} shows
that~$z_\precision$ can be computed using~$\cO(\precision
\order^{\omega-1})$ operations in~$\basefield$. We propose a solution
of better complexity.

By Cramer's rule,~$z$ is a vector of rational functions~$z_i(t)$, of
degree at most~$\order$.  The idea is to first compute~$z$ as a
rational function, and then to deduce its expansion
modulo~$t^{\precision +1}$. The first part of the algorithm does not
depend on~$\precision$ and thus it can be seen as a precomputation.
For instance, one can use%Algorithm \texttt{SeriesSolutionSmallRHS} in
~\cite[Corollary~12]{Storjohann02}, to compute $z$ in
complexity~$\cO(\order^{\omega} \log \order)$. In the second step of
the algorithm, we have to expand~$\order$ rational functions of degree
at most~$\order$ at precision~$\precision$.  Each such expansion
can be performed using~$\cO(\precision\sM(\order)/\order)$ operations
in~$\basefield$, see, e.g., the proof of~\cite[Prop.~1]{BoFlSaSc05}.
The total cost of the algorithm is thus~${\cO(\order^\omega\log \order
+ \precision \sM(\order))}$. We give below a simplified variant with
same complexity, avoiding the use of the algorithm
in~\cite{Storjohann02} for the precomputation step and relying instead
on a technique which is classical in the computation of minimal
polynomials~\cite{BuClSh97}.
\begin{enumerate}
\item Compute the vectors~$v,Av,A^2 v,A^3v,\dots,A^{2r}v$
  in~$\cO(\order^\omega\log \order)$, as follows: \\ for~$\kappa$
  from~$1$ to~${1 + \log \order}$ do
    \begin{enumerate}
    \item compute~$A^{2^\kappa}$
    \item compute~${A^{2^\kappa} \times [v | Av | \cdots | A^{2^\kappa-1}v]}$,
        thus getting~${[A^{2^\kappa}v | A^{2^\kappa+1}v | \cdots | A^{2^{\kappa+1}-1}v]}$
    \end{enumerate}
  \item For each~${j=1, \dots, \order}$:
\begin{enumerate}
\item recover the rational fraction whose series expansion
  is~$\sum{(A^i v)_j t^i}$ by Pad\'e approximation
  in~$\cO(\sM(\order)\log \order)$ operations
\item compute its expansion up to precision $t^{\precision + 1}$
  in~$\cO(\precision \, \sM(\order)/\order)$ operations
\item recover the expansion of~$y$ from that of~$z$,
  using~$\cO(\precision)$ operations.
\end{enumerate}
\end{enumerate}
This yields the announced total cost of ~${\cO(\order^\omega \log
\order + \precision \sM(\order))}$ operations for problem {\bf II}.

We now turn to the estimation of the
cost for problems~{\bf i} and~{\bf I} (bases of solutions). 
In the case of equations with constant coefficients, we use the
Laplace transform again. If $y = \sum_{i \geq 0} y_i t^i$ is a
solution of an order $\order$ equation with constant coefficients,
then the sequence $(z_i)=(i! y_i)$ is generated by a linear recurrence
with constant coefficients. Hence, the first terms $z_1,\dots,z_\precision$ can
be computed in $O(\precision\sM(\order)/\order)$ operations, using again the
algorithm described in~\cite[Prop.~1]{BoFlSaSc05}.
For problem~{\bf I}, the exponent~$\omega+1$ in the cost of the precomputation can be reduced to~$\omega$ by a very different approach; we cannot give the details here for space limitation.

%%%%%%%%%%%%%%%%%%%%%%%%%%%%%%%%%%%%%%%%%%%%%%%%%%%%%%%%%%%%

\subsection{Polynomial Coefficients}
If the coefficients in one of the problems {\bf i, ii, I}, and~{\bf II}
are polynomials in~$\basefield[t]$ of degree at most~$d$, using the
linear recurrence of order~$d$ satisfied by the coefficients of the
solution seemingly yields the lowest possible complexity.
Consider for instance problem~{\bf II}.
Plugging~${A=\sum_{i=0}^d t^i A_i}$, ${b=\sum_{i=0}^d t^i
  b_i}$, and~${y=\sum_{i\geq 0}^d t^i y_i}$ in the
equation~${y'=Ay+b}$, we arrive at the following recurrence
$$
y_{k+d+1} = (d+k+1)^{-1} (A_d y_k + A_{d-1}y_{k+1} + \dots + A_0
y_{k+d} + b_{k+d}), \quad \text{for all $k \geq -d$}.
$$
Thus, to compute~$y_0,\dots,y_\precision$, we need to
perform~${\precision d}$ matrix-vector products; this is done
using~${\cO (d \precision \order^2)}$ operations in~$\basefield$. A
similar analysis implies the other complexity estimates in the third
column of Table~\ref{table1}.

%%%%%%%%%%%%%%%%%%%%%%%%%%%%%%%%%%%%%%%%%%%%%%%%%%%%%%%%%%%%
%%%%%%%%%%%%%%%%%%%%%%%%%%%%%%%%%%%%%%%%%%%%%%%%%%%%%%%%%%%%
%%%%%%%%%%%%%%%%%%%%%%%%%%%%%%%%%%%%%%%%%%%%%%%%%%%%%%%%%%%%

\section{Non-linear Systems of Differential Equations}
Let~${\varphi(t,y) = (\varphi_1(t,y), \dots, \varphi_\order(t,y))}$,
where each~$\varphi_i$ is a power series
in~$\basefield[[t,y_1,\dots,y_\order]]$. We consider the first-order
non-linear system in~$y$
\[
%(\mathcal{N})\qquad \left\{
%\begin{aligned}
% y_1'(t)& = \varphi_1(t,y_1(t),\dots,y_\order(t)), \\
%&\,\,\vdots \\
% y_\order'(t)& = \varphi_\order(t,y_1(t),\dots,y_\order(t)).
%\end{aligned}
%\right.
(\mathcal{N})\qquad % \left\{
y_1'(t) = \varphi_1(t,y_1(t),\dots,y_\order(t)), \quad\dots,\quad
 y_\order'(t) = \varphi_\order(t,y_1(t),\dots,y_\order(t)).
%\right.
\]

To solve~($\mathcal{N}$), we use the classical technique of
\emph{linearization}. The idea is to attach, to an \emph{approximate}
solution~$y_0$ of~($\mathcal{N}$), a \emph{tangent} system in the new unknown~$z$,
$$
(\mathcal{T},y_0) \qquad z' = \Jac(\varphi)(y_0) z - y_0'+
\varphi(y_0),
$$
which is linear and whose solutions serve to obtain a
better approximation of a true solution of~($\mathcal{N}$).  Indeed,
let us denote by~$(\mathcal{N}_m),(\mathcal{T}_m)$ the
systems~$(\mathcal{N}),(\mathcal{T})$ where all the equalities are
taken modulo~$t^m$.  Taylor's formula states that the
expansion~${\varphi(y+z) - \varphi(y) - \Jac(\varphi)(y) z}$
is equal to~$0$ modulo~$z^2$.
It is a simple matter to check that if~$y$ is a
solution of~$(\mathcal{N}_m)$ and if~$z$ is a solution
of~$(\mathcal{T}_{2m},y)$, then~${y+z}$ is a solution
of~$(\mathcal{N}_{2m})$.  This justifies the correctness of 
Algorithm {\sf SolveNonLinearSys}.
 
 To analyze the complexity of this algorithm, it suffices to remark
 that for each integer~$\kappa$ between $1$ and~$\intpart{\log \precision}$,
 one has to compute one solution of a linear inhomogeneous first-order
 system at precision~$2^\kappa$ and to evaluate~$\varphi$ and its
 Jacobian on a series at the same precision. This concludes the proof of Theorem~\ref{theo:non-linear}.

\begin{figure}[h]
\begin{center} 
  \fbox{\begin{minipage}{9 cm}
      
      \medskip
      \begin{center}\textsf{SolveNonLinearSys}($\phi,v$) \end{center}
      \textbf{Input:} $\precision$ in~$\mathbb{N}$,
      $\varphi(t,y)$ in~$\basefield[[t,y_1,\dots,y_\order]]^{\order}$, 
      $v$ in~$\basefield^\order$
      \par\smallskip
      \textbf{Output:} first~$\precision$ terms of a~$y(t)$
      in~$\basefield[[t]]$ such that~${y(t)' = \varphi(t,y(t)) \mod
        t^\precision}$ and~${y(0) = v}$.
\begin{tabbing}
  \;\;    \\$m \leftarrow 1$\\
$y \leftarrow v$ \\
  \textsf{while} $m \leq \precision/2$ \textsf{do}\\
  \hspace{0.5cm} $A \leftarrow \trunch{\Jac(\varphi) (y)}{2m}$\\
  \hspace{0.5cm} $b \leftarrow \trunch{\varphi (y) - y'}{2m}$ \\
% \hspace{0.5cm} $z \leftarrow \textsf{Solve}(z' = Az + b \mod t^{2m}, z(0)=0)$ \\
  \hspace{0.5cm} $z \leftarrow \textsf{Solve}(A, b, 2m, 0)$ \\
  \hspace{0.5cm} $y \leftarrow y + z$ \\
  \hspace{0.5cm} $m \leftarrow 2m$ \\
  \textsf{return} $y$
\end{tabbing}
\end{minipage}
}\end{center}
\caption{Solving the non-linear differential system ${y' = \varphi(t,y), \; y(0) = v}$.}
\label{fig:nonlinear}
 \end{figure}

%%%%%%%%%%%%%%%%%%%%%%%%%%%%%%%%%%%%%%%%%%%%%%%%%%%%%%%%%%%%
%%%%%%%%%%%%%%%%%%%%%%%%%%%%%%%%%%%%%%%%%%%%%%%%%%%%%%%%%%%%
%%%%%%%%%%%%%%%%%%%%%%%%%%%%%%%%%%%%%%%%%%%%%%%%%%%%%%%%%%%%

 \section{Implementation and Timings}

 We implemented our algorithms \textsf{SolveDiffHomSys} and
 \textsf{Solve} in Magma~\cite{Magma} and ran the programs on an Athlon processor at 2.2~GHz
 with 2~GB of memory.\footnote{All the computations have been done on the machines of
   the MEDICIS ressource center
   \url{http://medicis.polytechnique.fr}.}  We used Magma's built-in
 polynomial arithmetic (using successively naive, Karatsuba and FFT
 multiplication algorithms), as well as Magma's scalar matrix
 multiplication (of cubic complexity in the ranges of our interest).
 We give three tables of timings. First, we compare in Figure
 \ref{fig:benchs} the performances of our algorithm
 \textsf{SolveDiffHomSys} with that of the naive quadratic algorithm,
 for computing a basis of (truncated power series) solutions of a
 homogeneous system.  The order of the system varies from $2$ to $16$,
 while the precision required for the solution varies from 256 to
 4096; the base field is $\mathbb{Z}/p\mathbb{Z}$, where $p$ is a 32-bit prime.

\begin{figure}
\begin{center}
\begin{tabular}{c||c|c|c|c}
$ \precision \ddots  \order$      &  2  & 4   & 8 & 16 \\
\hline 
256    &   0.02 \text{vs.}  2.09      & 0.08 \text{vs.}  6.11        &  0.44  \text{vs.}  28.16     &  2.859 \text{vs.}  168.96 \\
512    &  0.04 \text{vs.}  8.12       &  0.17 \text{vs.}  25.35     & 0.989  \text{vs.}  113.65  &  6.41 \text{vs.}  688.52 \\
1024  & 0.08 \text{vs.}  32.18      &  0.39  \text{vs.}  104.26  & 2.30 \text{vs.}  484.16     &  15   \text{vs.}  2795.71\\
2048  & 0.18  \text{vs.}  128.48   &  0.94 \text{vs.}  424.65   & 5.54 \text{vs.}  2025.68  &  36.62  \text{vs.} $> 3$\text{hours} $^\star$\\
4096  & 0.42 \text{vs.}  503.6      &  2.26 \text{vs.}  1686.42 &  13.69 \text{vs.}  8348.03  & 92.11  \text{vs.} $> 1/2$ \text{day}$^\star$ \\

\end{tabular}
\end{center}
\caption{Computation of a basis of a linear homogeneous system with
  $\order$ equations, at precision $\precision$: comparison of timings
  (in seconds) between algorithm \textsf{SolveDiffHomSys} and the
  naive algorithm. Entries marked with a `$\star$' are estimated timings.}
\label{fig:benchs}
\end{figure}

\begin{figure}[ht]
\begin{center}
\begin{minipage}[b]{0.45\textwidth}
\centerline{\includegraphics[scale=0.35,angle=270]{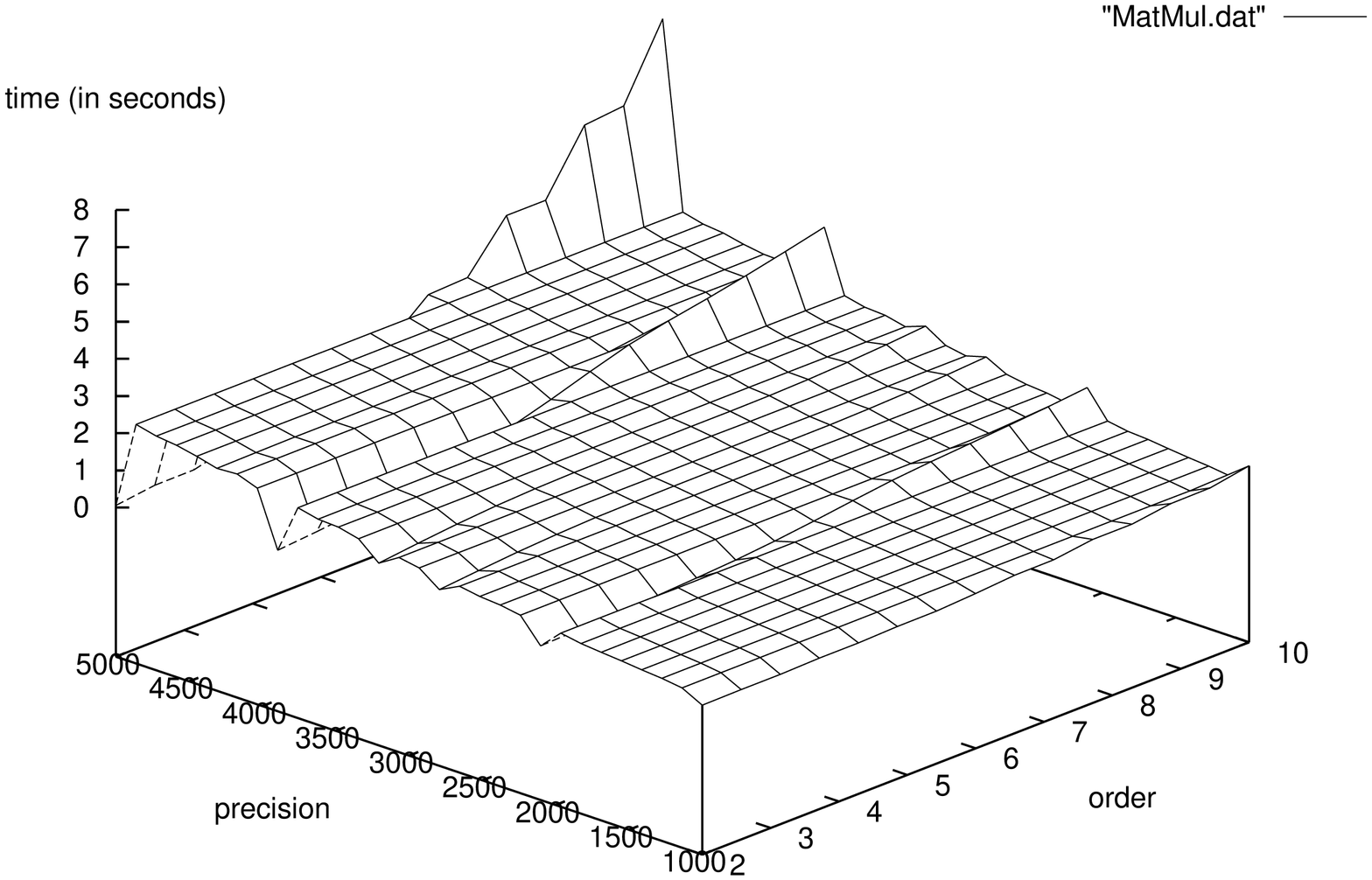}}
\caption{Timings of algorithm \textsf{PolyMatMul}.  \label{fig:matmul}}
\end{minipage}\hskip0.1\textwidth
\begin{minipage}[b]{0.45\textwidth}
\centerline{\includegraphics[scale=0.35,angle=270]{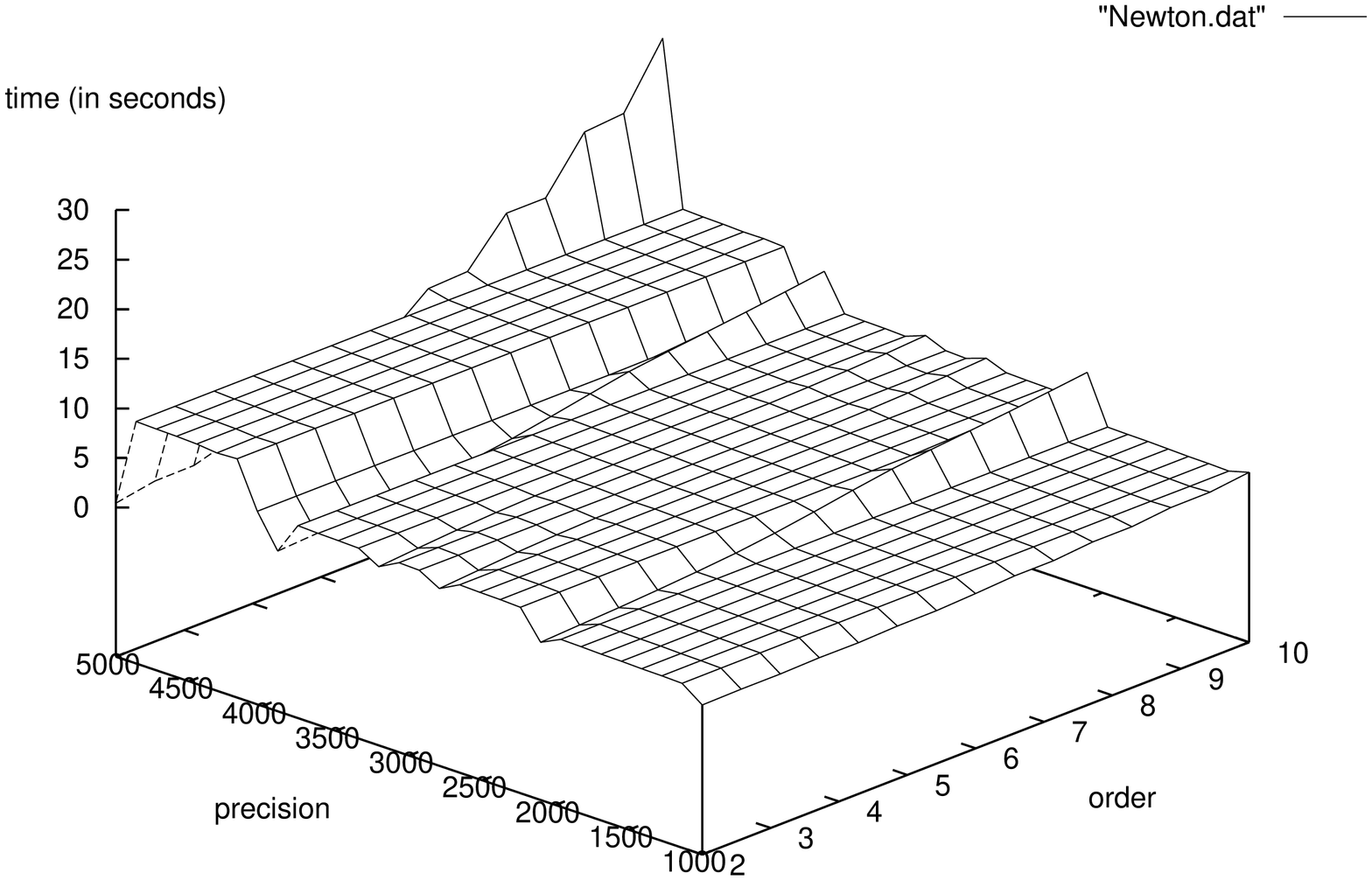}}
\caption{Timings of algorithm  \textsf{SolveDiffHomSys}.  \label{fig:newton}}
\end{minipage}
\end{center}
\end{figure}

Then we display in Figure~\ref{fig:matmul} and Figure~\ref{fig:newton}
the timings obtained respectively with
algorithm~\textsf{Solve\-DiffHomSys} and with the algorithm for
polynomial matrix multiplication \textsf{PolyMatMul} that was used as
a primitive of \textsf{SolveDiffHomSys}. The similar shapes of the two
surfaces indicate that the complexity prediction of point (a) in
Theorem~\ref{theo:linear} is well respected in our implementation:
\textsf{SolveDiffHomSys} uses a constant number (between 4 and 5) of
polynomial multiplications; note that the abrupt jumps at powers of 2
reflect the performance of Magma's FFT implementation of polynomial
arithmetic.

\begin{figure}[ht]
\begin{center}
\begin{minipage}[b]{0.45\textwidth}
\centerline{\includegraphics[scale=0.3,angle=270]{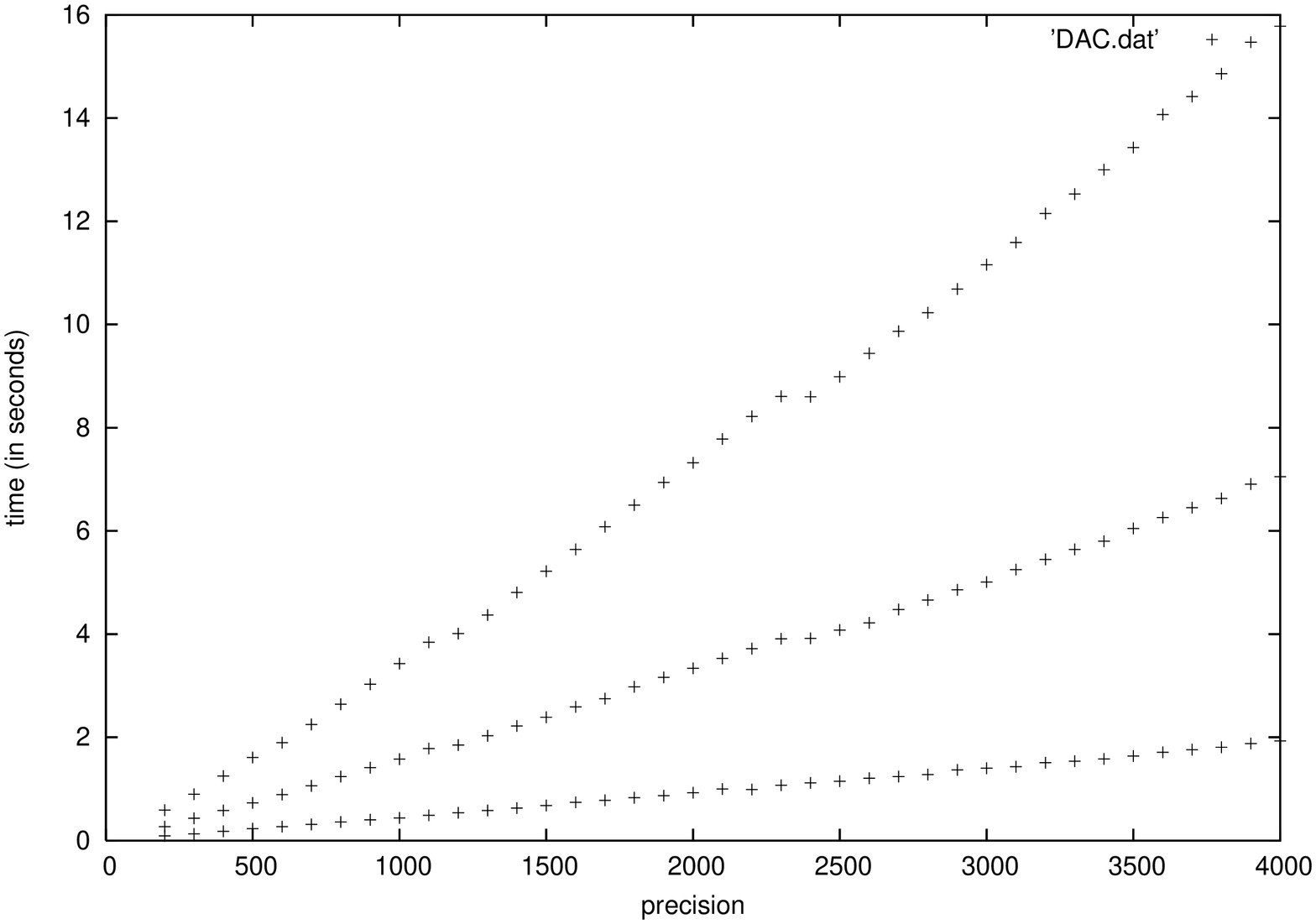}}
\caption{Timings of algorithm \textsf{Solve} for equations of orders 2, 4, and~8. \label{fig:dac}}
\end{minipage}\hskip0.1\textwidth
\begin{minipage}[b]{0.45\textwidth}
\centerline{\includegraphics[scale=0.3,angle=270]{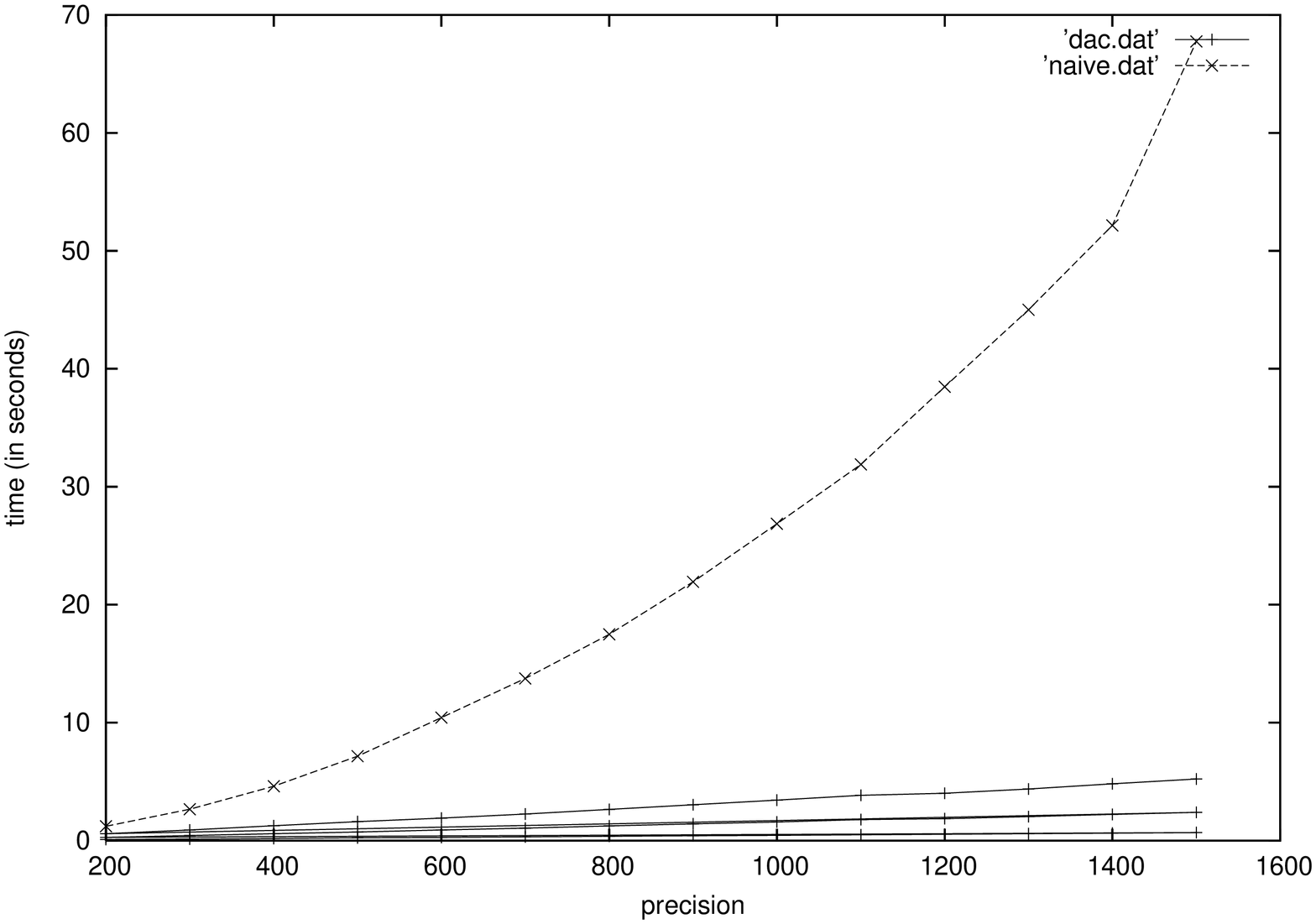}}
\caption{Same, compared to the naive algorithm for a second-order equation.\label{fig:dac+naive}}
\end{minipage}
\end{center}
\end{figure}

In Figure ~\ref{fig:dac} we give the timings for the computation of
one solution of a linear differential equation of order $2$, $4$, and
$8$, respectively, using our algorithm~\textsf{Solve} in
Section~\ref{sec:DAC}. Again, the shape of the three curves
experimentally confirms the nearly linear behaviour established in
point (b) of Theorem~\ref{theo:linear}, both in the
precision~$\precision$ and in the order $\order$ of the complexity of
algorithm ~\textsf{Solve}. Finally, Figure~\ref{fig:dac+naive}
displays the three curves from Figure~\ref{fig:dac} together with the
timings curve for the naive quadratic algorithm computing one solution
of a linear differential equation of order $2$.  The conclusion is
that our algorithm~\textsf{Solve} becomes very early superior to the
quadratic one.

%precision~$\precision = 1048576$ in~24.53s; one at doubled
%precision~$\precision=2097152$ in doubled time~49.05s; one for doubled
We also implemented our algorithms of Section~\ref{ssec:const-coeffs}
for the special case of constant coefficients. For reasons of space
limitation, we only provide a few experimental results for
problem~{\bf II}.  Over the same finite field, we computed: a solution
of a linear system with~$\order=8$ at
precision~$\precision\approx10^6$ in~24.53s; one at doubled precision
in doubled time~49.05s; one for doubled order~$\order=16$ in doubled
time~49.79s.

\bibliographystyle{plain}
\bibliography{focs}

\end{document}